\documentclass[useAMS,usenatbib]{mn2e}
\usepackage{epsfig,amsmath,natbib}
\usepackage{color}

\newcommand{\xhi}{$x_{\rm{HI}}$\,}

\def\prosima{$\; \buildrel \propto \over \sim \;$}
\def\be{\begin{equation}} 
\def\ee{\end{equation}}

\def\msun{{M_\odot}} 
  
\def\rvir{r_{vir}}

\def\HI{\hbox{H~$\scriptstyle\rm I\ $}}
\def\xhi{$x_{\rm HI}$}

\def\NHI{{N_{\rm HI}}}  
 
\def\gsim{\lower.5ex\hbox{\gtsima}} 
\def\lsim{\lower.5ex\hbox{\ltsima}} \def\gtsima{$\; \buildrel > \over 
\sim \;$} \def\ltsima{$\; \buildrel < \over \sim \;$} \def\prosima{$\; 
\buildrel \propto \over \sim \;$} \def\gsim{\lower.5ex\hbox{\gtsima}} 
\def\lsim{\lower.5ex\hbox{\ltsima}} 
\def\simgt{\lower.5ex\hbox{\gtsima}} 
\def\simlt{\lower.5ex\hbox{\ltsima}} 
\def\simpr{\lower.5ex\hbox{\prosima}} \def\la{\lsim} \def\ga{\gsim}

\def\gtsima{$\; \buildrel > \over \sim \;$} 
\def\ltsima{$\; \buildrel < \over \sim \;$} 
\def\gsim{\lower.5ex\hbox{\gtsima}} 
\def\lsim{\lower.5ex\hbox{\ltsima}} 
\def\simgt{\lower.5ex\hbox{\gtsima}} 
\def\simlt{\lower.5ex\hbox{\ltsima}} 
\def\simpr{\lower.5ex\hbox{\prosima}} 
\def\la{\lsim} 
\def\ga{\gsim}

\def\Lya{Ly$\alpha$~}

\def\msun{\,{\rm M_\odot}}

\def\E3{{\cal E}_{\rm g}^{III}}

\title[LAEs powered by AGN]{Identifying Lyman Alpha Emitters powered by AGNs}         
\author[S. Baek, A. Ferrara]
{Sunghye Baek\thanks{E-mail: sunghye.baek@sns.it}, Andrea Ferrara\\
Scuola Normale Superiore, Piazza dei Cavalieri 7, 56126 Pisa, Italy}

\date{} 
\begin{document}
\maketitle
\label{firstpage} 
\begin{abstract} 
Lyman Alpha Emitters (LAEs) are usually thought to be powered by star formation. It has been recently reported that a fraction of LAEs at redshift $z\sim3-4$ hosts an Active Galactic Nuclei (AGN). If an AGN is present it could be obscured and undetectable in X-rays, but yet dominate the Ly$\alpha$ luminosity. We examine the properties of these AGN-powered LAEs at high redshift ($z\ge6$) using radiative transfer cosmological simulations and obtain a reliable criterion to identify them from their observed \Lya line and surface brightness. We find that these sources should have: (a) \textit{negative} line weighted skewness, $S_w <0$, and (b) surface brightness profiles FWHM $\ge1.5"$. This parameter space cannot be populated by starburst LAEs. Thus, LAEs satisfying this criterion would be strong candidates for the presence of a hidden AGN powering their luminosity.
\end{abstract}


\section{Introduction}
Lyman Alpha Emitters (LAEs) are galaxies emitting a prominent \Lya $1216 \rm{\AA}$  emission line. Narrow band imaging techniques have allowed the discovery of thousands of LAE candidates in an extended redshift range $z\simeq 2-7$ \citep{cowi98, Malh04, Shim05, Ouch05, Ouch08, Kash06} and hundreds of them have been spectroscopically confirmed \citep{Hu04, Daws04, Ouch08, Kash06}. 

While the most natural explanation for the large Ly$\alpha$ luminosities is associated with a star forming activity, interestingly, a fraction of LAEs seem to host an Active Galactic Nucleus (AGN). This important property is now confirmed by X-ray detections reported by  \citet{Gawi06} at $z\sim3$, and by \citet{Zhen10b} at $z\sim4.5$. 
The source J033127.2-274247 discovered by \citet{Zhen10b} shows a strong soft band ($0.5-2$ keV) X-ray flux with $L_X=4.2\times 10^{44}\,\rm{erg\,s^{-1}}$, 
and has been spectroscopically confirmed as an unobscured AGN. 
\citet{Ouch08} found broad emission lines -- a typical AGN signature -- among photometrically selected LAEs at $z=3.1$ and $z=3.7$. 
The fraction of LAEs showing broad emission line is small ($\approx 1$\%); however, the brightest LAEs ($L_\alpha\gsim 10^{43.4}\rm{erg\,s^{-1}}$) 
always appear to host an AGN. 

Unfortunately, at high redshifts, where assessing the presence of a central accreting black hole would be of outmost importance, e.g. to study the precursors of the super-massive black holes, X-ray fluxes are too faint to be used as a discriminating tool. In addition, it is likely that in the majority of these sources X-ray emission from the central black hole is obscured by a dense absorbing gas layer. One would then like to be able to infer the presence of an AGN directly from the properties of the Ly$\alpha$ emission, which most often is the only information available for these distant sources.  
The giant LAE  at $z=6.6$ named Himiko \citep{Ouch09b} is suspected to contain a hidden AGN due to its prominent size and Ly$\alpha$ luminosity. The lack of detection in either X-ray, MIR, sub-mm and radio bands are not sufficient to discard such possibility. In this work, based on high resolution, radiative transfer (RT) cosmological simulations of a LAE similar to Himiko, we 
derive a criterion to unambiguously assess if a generic LAE is powered by an AGN or by a starburst.

\section{Numerical simulations}
\begin{figure}
\begin{center}
 \includegraphics[width=60mm]{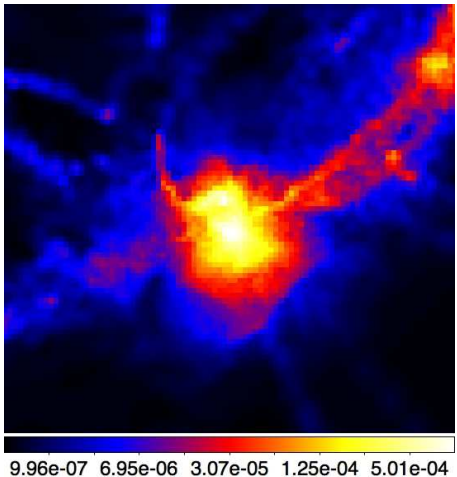}
 \end{center}
 \caption{Gas density field  in a slice of thickness 1 kpc and proper size 120 kpc, centered on the most massive halo with $M_h=1.17\times10^{11}\msun$ at $z=6.6$. The colorbar shows the density values in [$\rm{cm^{-3}}$]. } 
 \label{Mh_sfr}
\end{figure}

We start by running a cosmological SPH hydrodynamic simulation\footnote{We use the recent WMAP7+BAO+$H_0$ cosmological parameters: $\Omega_m=0.272$, $\Omega_{\Lambda}=0.728$, $\Omega_b=0.0455$, $h=0.704$, $\sigma_8=0.807$ \citep{Koma11}.} using GADGET-2 \citep{Spri05} and extract a snapshot
at $z=6.6$. We simulate a $(10h^{-1}\rm{Mpc})^3$ volume with $2\times512^3$ baryonic+dark matter particles, giving a mass resolution of $(1.32, 6.68)\times10^{5}\msun$ for (baryons, dark matter).
We post-process the snapshot by running UV and X-ray RT using LICORICE \citep{Baek09,Baek10}, so that 
the remaining average \HI fraction, \xhi, in the intergalactic medium (IGM) is 0.1, in agreement with 
current data \citep{Mort11}.  By using a friends-of-friends algorithm we identify the most massive halo, of total mass $M_h=1.17\times10^{11}\msun$,
gas mass $M_g=2\times10^{10}\msun$, and virial radius $\rvir \approx20$ proper kpc (pkpc). Fig.\ref{Mh_sfr} shows the gas density map of the halo, connected to the filamentary dense structure. 
Then we select a $(0.625\,h^{-1}\rm{Mpc})^3$ comoving volume centered on the halo and re-simulate the UV/X-ray RT at higher spatial resolution on an adaptive grid  with $N_{max}=8$ according to the \citet{Baek09} scheme\footnote{The $N_{max}=8$ value ensures that the RT cell contains only few particles:  of the total number of cells, a fraction (0.31, 0.28,0.16) contain (1,2,3) particles. The number of UV photons in a photon packet has to be smaller than the number of H atoms in the cell. Even with adaptive integration time steps, this is computationally too demanding. Therefore gas particles outside the re-simulated box are set to have \xhi=0.1 and $T=10^4$ K for simplicity.}, giving a minimum RT cell size of 0.114 pkpc.

We have run simulations for three different Spectral Energy Distributions (SED) of the central source, namely (a) starburst, 
(b) Compton-thick, (c) Compton thin AGN types, keeping all other conditions the same, and setting the 
ionizing photons escape fraction, $f_{esc}=0$. For the starburst case, radiation is emitted isotropically with a bolometric luminosity obtained from STARBURST99 \citep{Leit99} assuming a continuous star formation rate of $10\msun \rm{yr}^{-1}$, metallicity $Z=10^{-4}$, a Salpeter initial mass function $\alpha=2.35$ in the mass range $1-100 \msun$. The resulting photoionization rate is $Q=3\times 10^{54}\,\rm{s^{-1}}$.  For AGNs we use the composite spectra from radio to X-rays following \citet{Shan11}, with an X-ray upper cut-off\footnote{More energetic photons have a mean free path larger than the galaxy size and hence do not contribute to \Lya emissivity}  at 0.5 keV; the spectrum is normalized so that all cases have the above photoionization rate $Q$. Compton-thick AGNs emit their ionizing radiation ($h\nu>13.6$ eV) in a cone with opening angle $\theta =45^{\circ}$ with respect to the polar axis perpendicular to the line of sight; for Compton-thin AGNs, in addition, energetic X-ray photons ($h\nu>0.1$ keV) are allowed to emerge from the dusty torus (size $\approx$ 1 pc) and propagate isotropically into the ISM. The rest-frame AGN X-ray luminosity\footnote{These sources are then 10 times fainter than the detection limit of XMM-{\it Newton} X-ray in the same band.} in the energy band corresponding to the 0.5-2 keV observed band is $2.42\times10^{43}\,\rm{erg\,s^{-1}}$.

In addition to H recombinations, Ly$\alpha$ photons can be produced also by X-rays via de-excitations. Soft X-rays (0.1-0.5 keV) have a relatively short ($1-100$ kpc)  mean free path in the interstellar medium of $z\gsim 6$ galaxies. Once the X-ray photon is absorbed by an H atom, the primary electron of energy $E_p= h\nu-13.6$ eV can produce further (secondary) ionizations, excitations and 
heating of the gas; de-excitations from upper energy levels to the ground state eventually release Ly$\alpha$ 
photons. The fractional energy of the primary electron that goes into Ly$\alpha$ photons \citep{Vald08} depends on \xhi: if \xhi $> 0.99$, $f^X_{\alpha}(x_{\rm{HI}}) \approx 30$\%; such efficiency decreases to $f^X_{\alpha}(x_{\rm{HI}})=0.01$ for \xhi = 0.1.

Following X-ray photon absorption we compute the additional production of Ly$\alpha$ photons by de-excitations as 
\be
\Delta N_{\rm{Ly}\alpha}=N_{abs}\frac{E_p}{h\nu_{\alpha}}f^X_{\alpha}(x_{\rm{HII}}),
\ee
where $N_{abs}$ is the absorbed number of photons and $h\nu_{\alpha}$ the Ly$\alpha$ photon energy. During UV/X-ray RT, we store the number of recombination and de-excitation Ly$\alpha$ photons produced by each SPH particle.  
 
Finally Ly$\alpha$ RT is run using the LICORICE Ly$\alpha$ module \citep{Seme07} on a volume of $0.625\times0.625\times 10 \, h^{-3}\rm{Mpc}^3$. The size in the sky plane is sufficiently large ($20" \times 20"$) to model the 
surface brightness (SB) map of a LAE; the larger size (240 bins with rest frame size 0.1 \AA, centered on $\lambda _{\alpha}$) along the line of sight/frequency axis is required to properly account for the \Lya IGM transmissivity. We interpolate all gas physical properties on a fixed $128\times 128\times 2048$ grid using the SPH kernel, thus obtaining a spatial resolution of 0.918 pkpc. This is almost identical to the pixel scale achieved by the Subaru/Suprime-Cam \citep{Miya02} in the Subaru/XMM-Newton Deep Sky Survey \citep{Ouch08}.
We compute the Ly$\alpha$ luminosity from each cell, $L^{cell}_{\alpha}$, by summing all Ly$\alpha$ photons emitted by the SPH particles in the cell,
\be
L^{cell}_{\alpha}=\frac{1}{\Delta t}\sum_{i=1}^{N_{cell}}N_{\rm{Ly}\alpha}^i,
\ee
where $N_{cell}$ is the number of SPH particles in the cell and $i$ is the particle index. If $L^{cell}_{\alpha} \ge 10^{37} \rm{erg\,s^{-1}}$, the cell is considered as a Ly$\alpha$ source\footnote{Even for a detection limit as low as $10^{-21} \rm{erg\,s^{-1} cm^{-2} arcsec^{-2}} $, cells with $L^{cell}_{\alpha} < 10^{37} \rm{erg\,s^{-1}}$ contribute negligibly.},  and we cast Ly$\alpha$ photon packets from it.

\section{Results}

\begin{figure}
 \includegraphics[width=80mm]{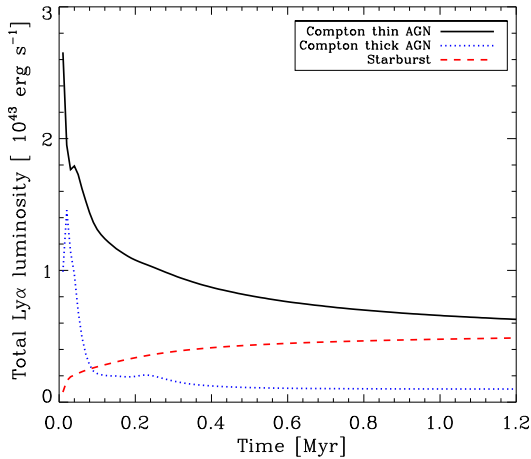}
 \caption{Time evolution of the total Ly$\alpha$ luminosity of the simulation box. The various curves refer to starburst (red dashed line), Compton-thin (black solid) and Compton-thick (black dotted) SED cases. 
} 
 \label{total_lya}
\end{figure}

We show the total (recombination + deexcitation) Ly$\alpha$ luminosity, $L_\alpha$, of the simulation box as a function of time in Fig.  \ref{total_lya}. For the starburst case $L_\alpha$ increases rapidly as the gas gets progressively ionized by the central source, approaching an asymptotic value of $5\times 10^{42}$ erg s$^{-1}$, set by the equilibrium between recombination and photoionization rates. Breakout of the ionization front (IF, defined by the position where as  $x_{\rm{HII}}=0.5$) from the galaxy occurs at 0.25 Myr. 

For the Compton-thick AGN the evolution is different.  Initially UV photons are absorbed locally, but X-rays penetrate to larger distances thus making the IF much smoother than for the starburst SED; X-rays absorbed beyond the IF effectively produce additional Ly$\alpha$ photons. The IF also propagates faster than in the starburst case, as the same photoionizing rate $Q$ is now confined in a cone.  While the IF travels in the ISM, X-rays boost the Ly$\alpha$ luminosity by several times with respect to the starburst SED; however, beyond the halo virial radius the low IGM density strongly reduces the \Lya emission;  hence after 0.1 Myr the total Ly$\alpha$ luminosity drops as the ionized volume where recombinations occur is confined to a cone. Thus we can identify two separate regimes for Ly$\alpha$ luminosity, namely (a) an X-ray dominated regime (IF in the ISM),  and (b) a recombination dominated regime (IF in the IGM).

For the Compton-thin AGN, in which X-rays are emitted isotropically, de-excitations in the neutral ISM outside the torus can largely amplify the Ly$\alpha$ luminosity with respect to the Compton-thick case as a larger fractional energy of X-rays can be converted into Ly$\alpha$ photons. Therefore, the case of Compton-thin AGN produces the highest luminosity among all cases, and converges slowly than the Compton-thick AGN case.

\begin{figure*}
 \includegraphics[width=45mm]{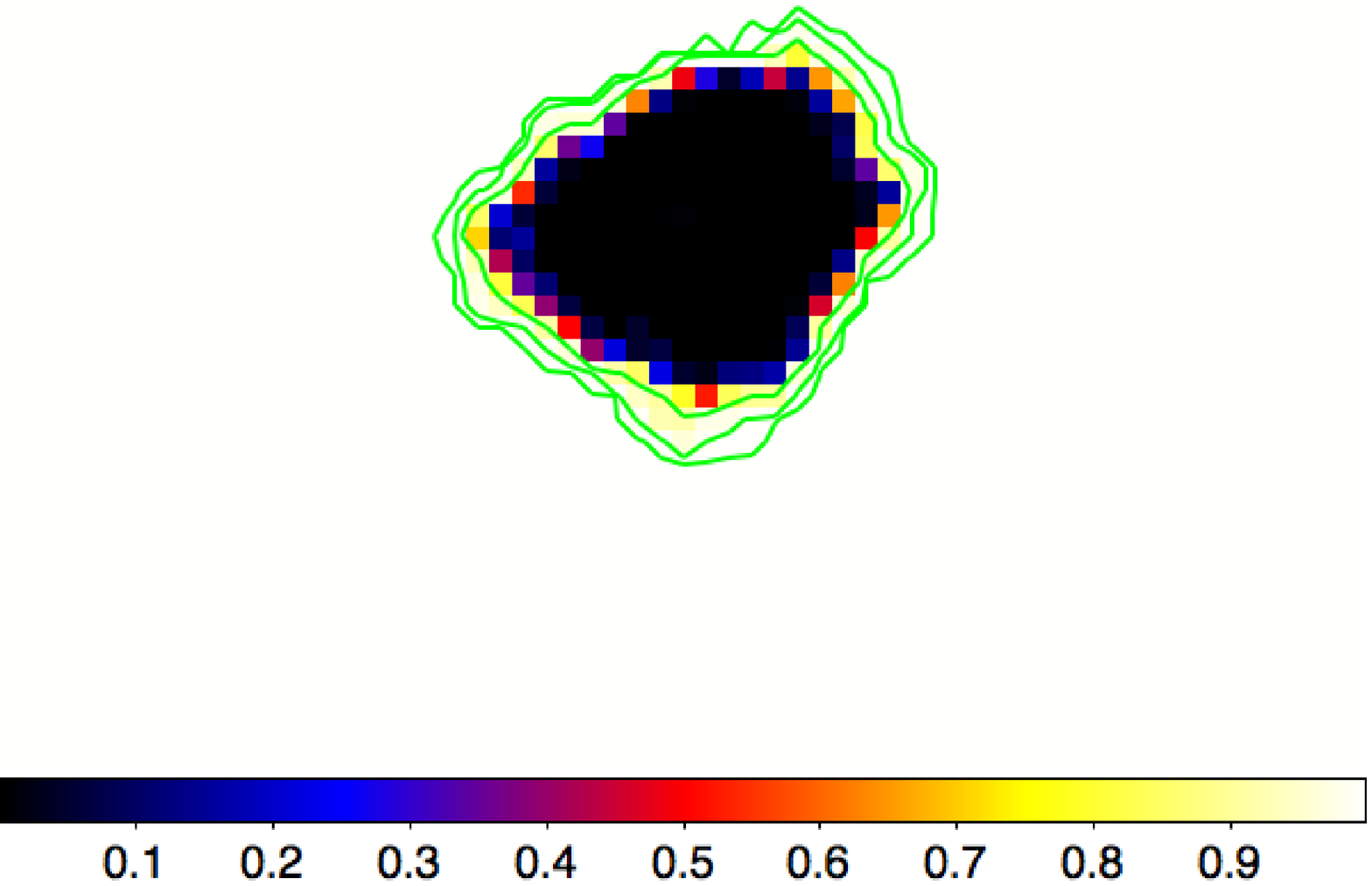}
 \includegraphics[width=45mm]{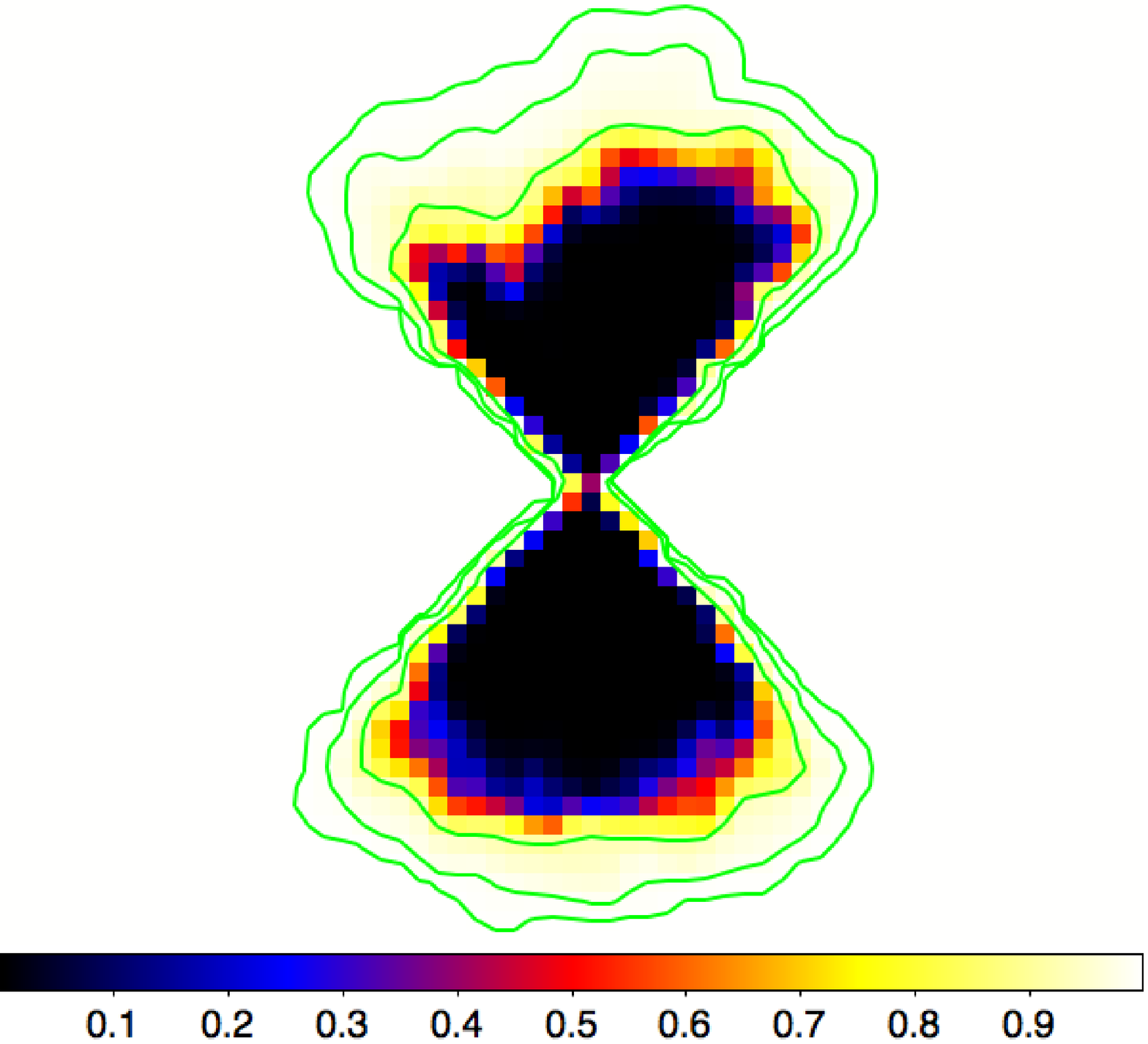}
 \includegraphics[width=45mm]{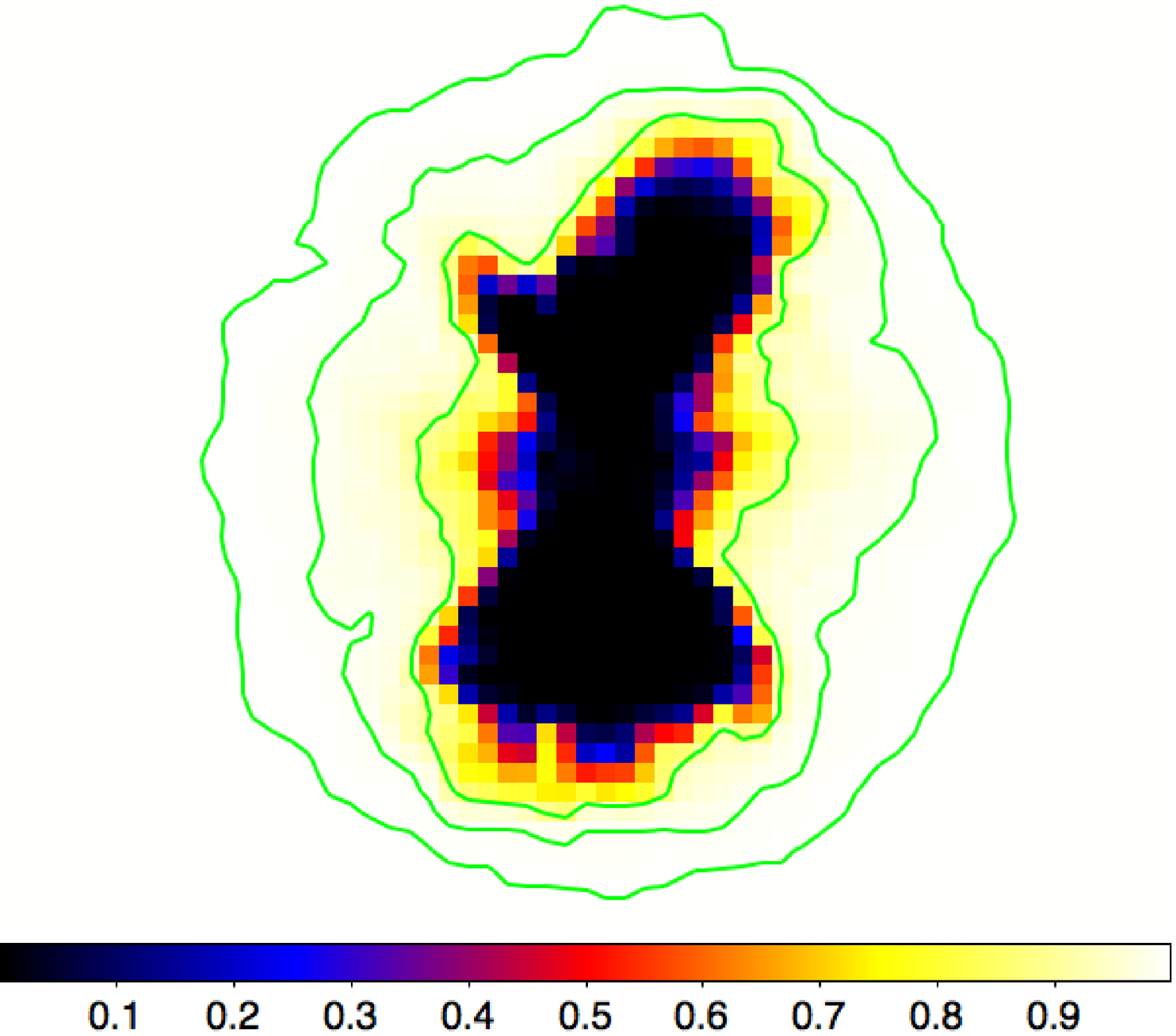}
  
  \includegraphics[width=45mm]{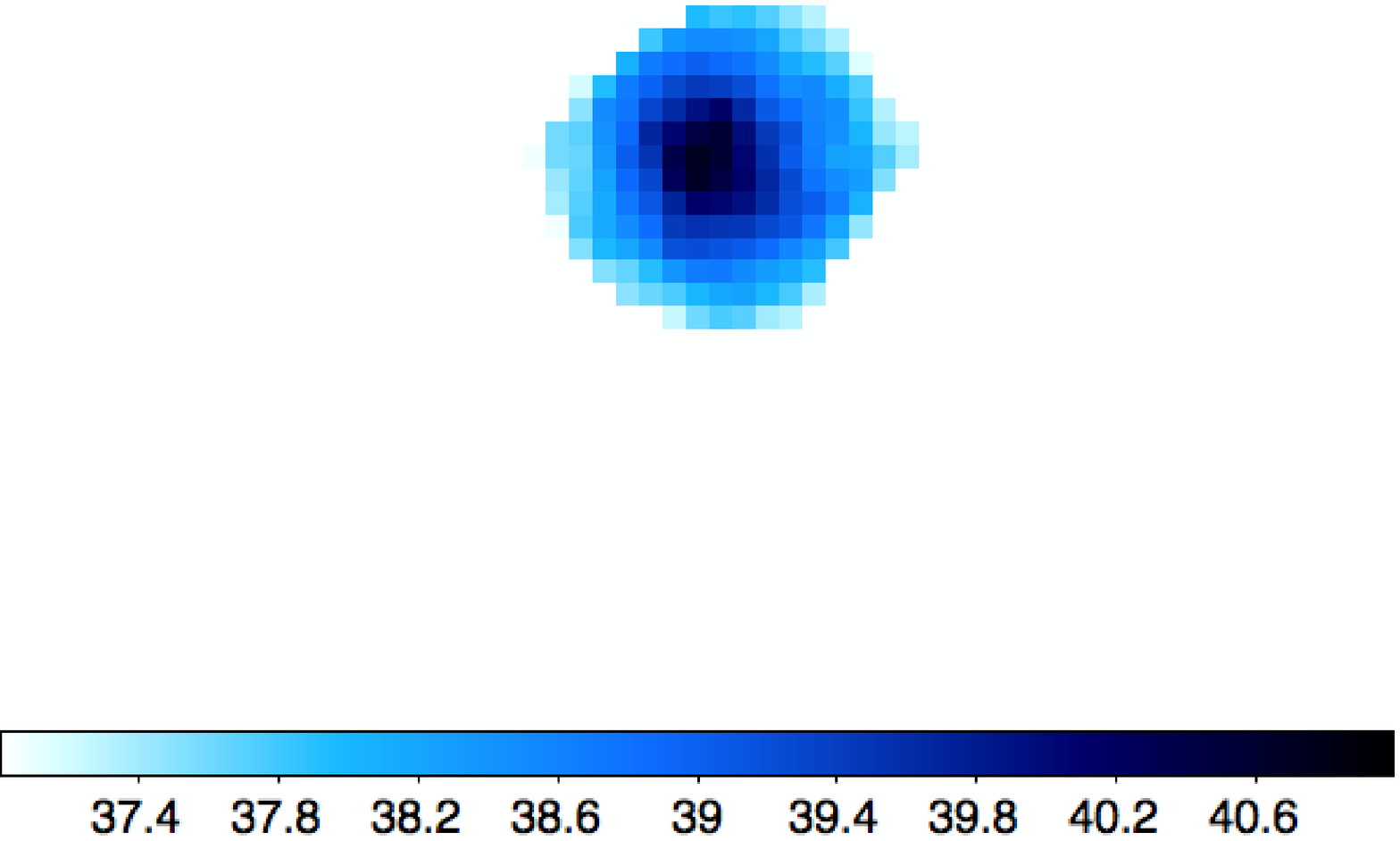}
  \includegraphics[width=45mm]{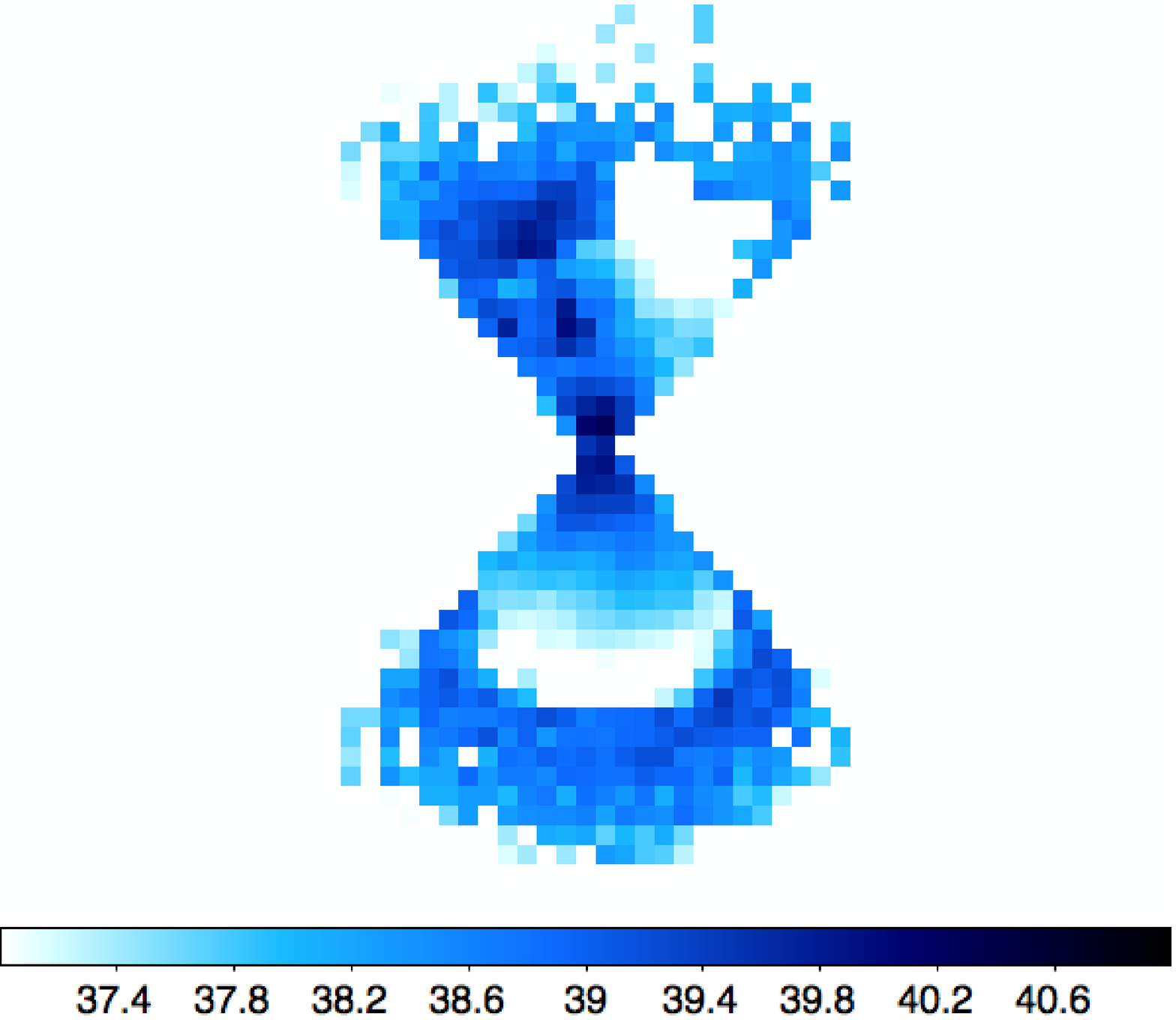}
  \includegraphics[width=45mm]{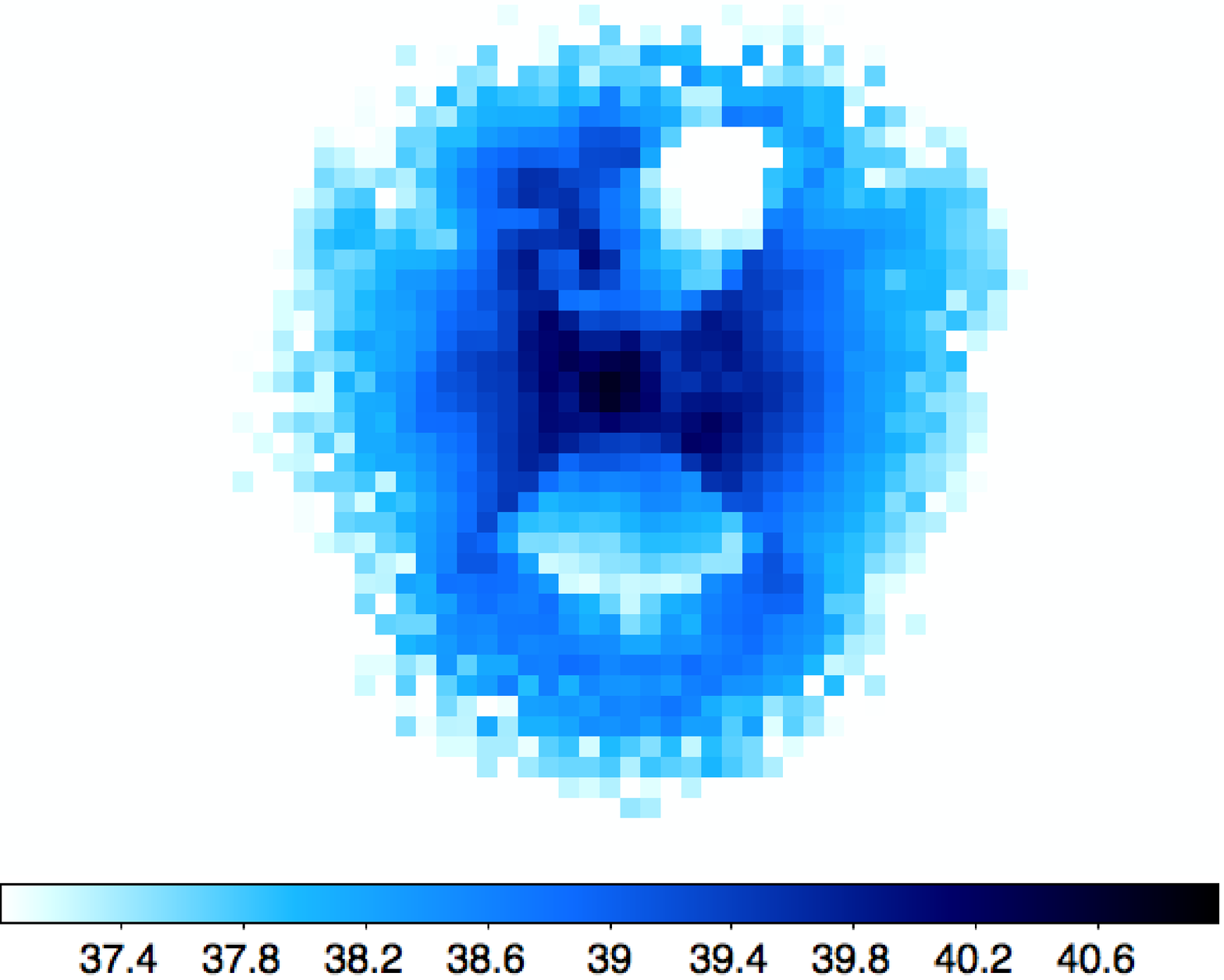}

 \includegraphics[width=45mm]{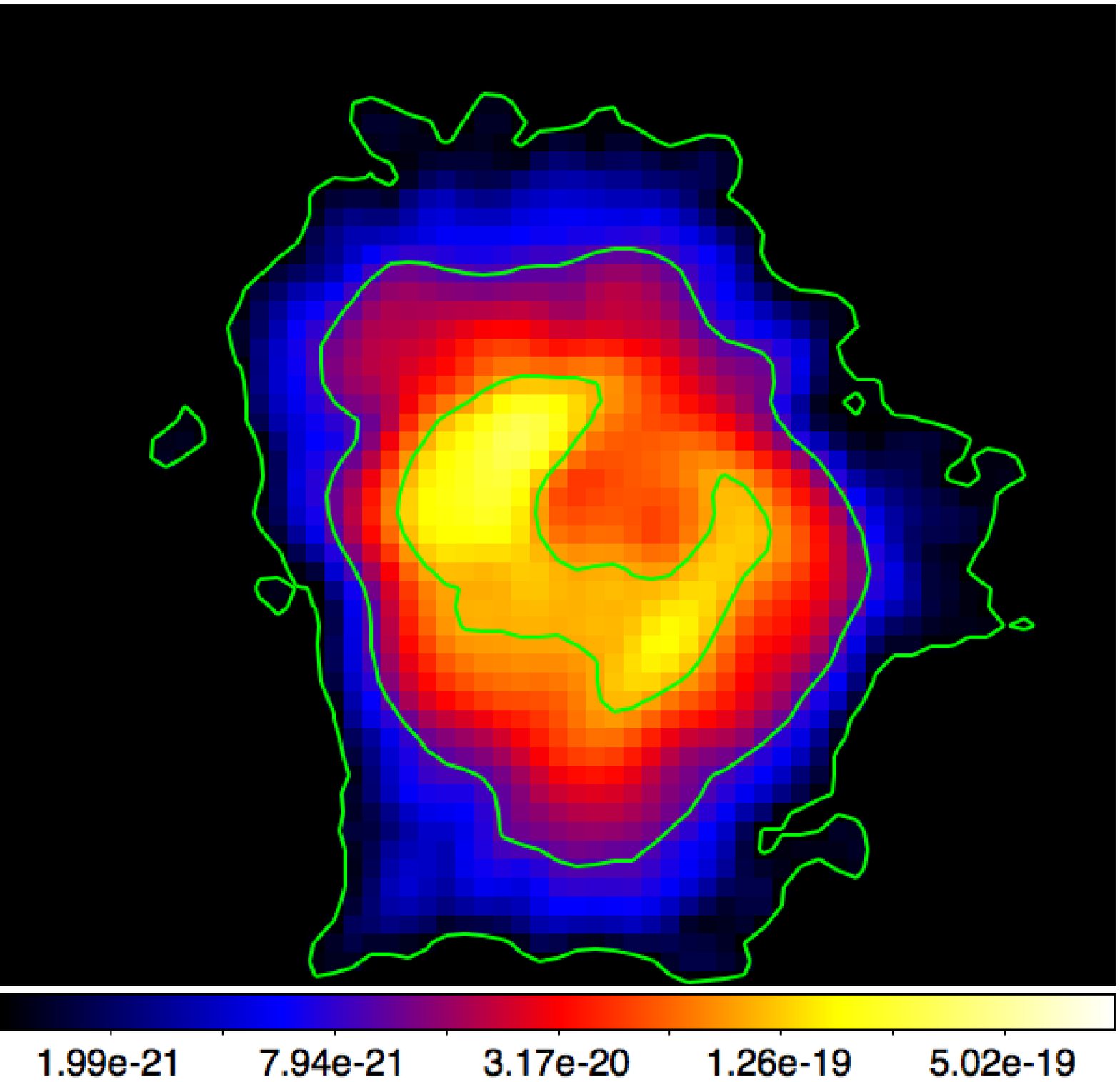}
 \includegraphics[width=45mm]{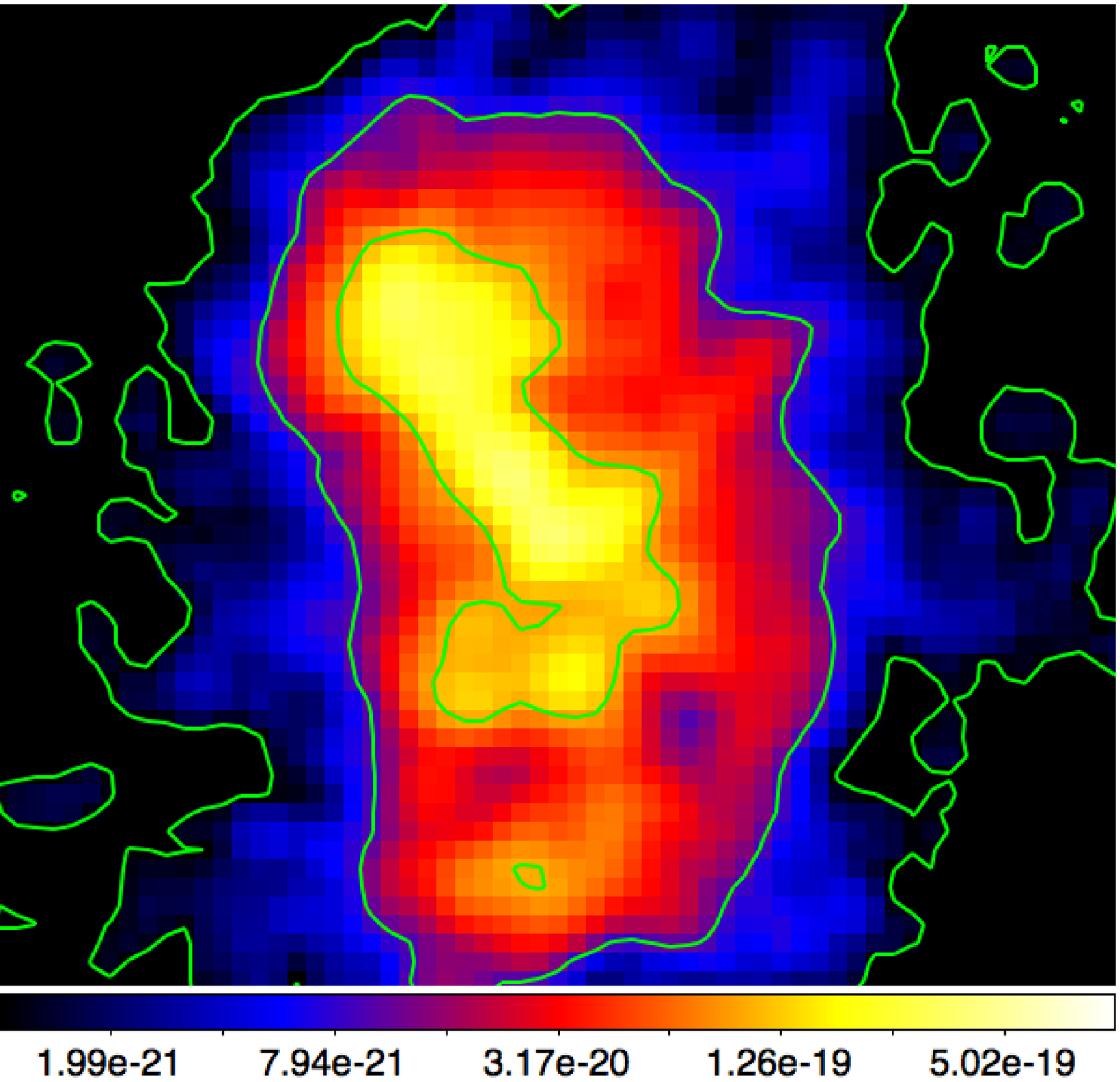}
 \includegraphics[width=45mm]{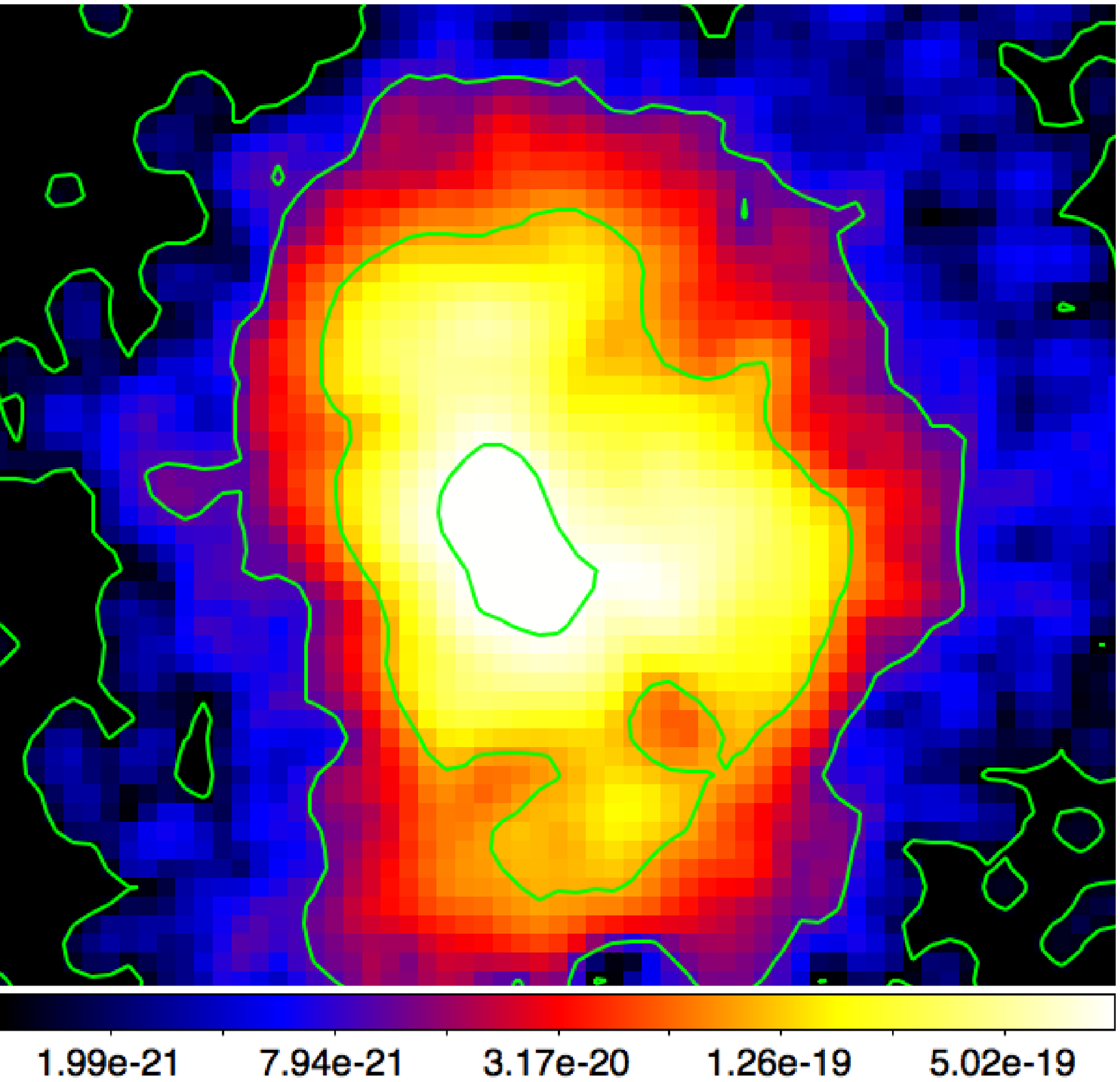}

 \caption{Maps of the \HI fraction with contour levels \xhi =0.9, 0.99, 0.999 (top row),  Lya emissivity (middle), and SB with contour levels $10^{-21+j}, j=0,..,3\rm{erg\,s^{-1}\,cm^{-2}\,arcsec^{-2}}$, smoothed with a gaussian of FWHM=0.8" (bottom) for the (a) starburst (left column), (b) Compton-thick AGN (middle), and (c) Compton-thin AGN (right) cases.  Snapshots are taken 0.06 Myr after source turn-on. Maps have size 60$\times$60 pkpc and are 3 pkpc-thick through the center of the most massive halo in the box.} 
 \label{panel}
\end{figure*}

X-rays produce Ly$\alpha$ very efficiently in the initial phases of photoionization, when the ISM is largely neutral. This can be appreciated from Fig. \ref{panel}, showing ionization fraction, Ly$\alpha$ luminosity, and  SB maps extracted 0.06 Myr after source turn-on for the three source cases. The snapshot at 0.06 Myr shows the most distinctive feature between starburst powered and AGN powered LAEs. The starburst SED produces a very sharp IF as expected, while the AGN SEDs are characterized by smoother IFs. For the Compton-thin case, the two outer contours (\xhi = 0.99, 0.999) are quite isotropic; nevertheless, the cone-shaped HII region is clearly visible. 

In the starburst case, Ly$\alpha$ photons are produced only inside the HII region through recombinations. As the 
recombination rate is proportional to $n^2$, the Ly$\alpha$ luminosity map closely resembles the density distribution, i.e. it 
is brighter at the center; this property also holds for the two AGN cases within the cone. Apart from these dense, ionized regions where recombination-driven Ly$\alpha$ production dominates, Ly$\alpha$ are created by X-rays also near and beyond the IF. As Compton-thin AGNs emit X-rays ($h\nu>100$ eV) isotropically, a copious amount of Ly$\alpha$ photons comes from regions shielded from the UV flux. 

As for the SB distribution (Fig. \ref{panel}, bottom row), AGNs show longitudinally extended shapes, while the starburst case is more isotropic. Even though Ly$\alpha$ emission is stronger in the central regions of the galaxy, SB maps do not exactly match the corresponding $L_\alpha$ maps.  This is due to the fact that most Ly$\alpha$ photons experience considerable frequency and spatial shifts through repeated scatterings with \HI atoms in the ISM. After being scattered $\approx 10^5$ times, the photon frequency is shifted away from the Ly$\alpha$ line center, allowing escape even if the \HI column density is as high as $\NHI = 10^{21}\,\rm{cm^{-2}}$. 
None of pixels in the starburst and Compton-thick AGN cases is above the current detection limit of narrow band images, $10^{-18}$  $\rm{erg\,s^{-1}\,cm^{-2}\,arcsec^{-2}}$; for the Compton-thin AGN instead, the region within the central 1.5 $\rm{arcsec^2}$ can be detected. The halo is connected with dense filamentary structure as shown in Fig.\ref{Mh_sfr}, but we checked that the total amount of HI in the filament is not enough to produce detectable Ly$\alpha$ emission. 

\subsection{Skewness evolution}

From the observed emission \Lya line profile we can compute the line skewness, $S$,  
\be
S=\frac{1}{I\sigma ^3}\int ^n_{i}(x_i-\bar{x})^3 f_i,
\ee
where $f_i$ is a 2D of the flux, $x_i$ is the pixel coordinate, $I=\int ^n_i f_i$, and $\bar{x}$, $\sigma$ are the 
mean and dispersion of $x_i$. Due to its higher sensitivity to profile asymmetries, the weighted skewness, 
$S_w=S(\lambda_{10,r}-\lambda_{10,b})$, where $\lambda_{10,r}$ and $\lambda_{10,b}$ are the wavelengths where the flux drops to 10\%
of the emission peak value on the red and blue sides of the line \citep{Kash06}, is more often used. 
Typically the \Lya line from high-$z$ sources,  differently from other emission lines such as [O II] or [O III] at 
lower redshifts, has $S_w>0$. Thus, $S_w$ can be an useful tracer of high-$z$ objects. 

For each SED case, we compute $S_w$ for 5 snapshots ($t=0.02, 0.04, 0.06, 0.08, 1$ Myr),
the first four (last) corresponding to the X-ray (recombination) dominated regime.
It is well known that for a spherical, static case \citep{Neuf90,Dijk06}, the emerging line profile 
shows a double peak shape\footnote{The size of the core is  
$x = {(\nu-\nu_\alpha)}/{\nu_D}\approx3$ where $\nu_D=({v_{th}}/{c})\nu_{\alpha}$ is the Doppler frequency shift and $v_{th}$ is thermal velocity dispersion of the gas.}, with the peaks getting further apart as $N_{\rm{HI}}$ is increased. The blue wing is then suppressed as photons propagate through the IGM if the source is located at sufficiently high redshift $z\gsim 5$; the red wing can be instead transmitted to the observer, giving $S_w >0$. 

During the X-ray dominated stage of Ly$\alpha$ production $\NHI$ is very high ($10^{18-20}\rm{cm^{-2}}$), corresponding to an optical depth at line center $\tau_\alpha =10^{6-8}$. The peak of the red wing is shifted farther to the red and the profile decline on the blue part of the peak becomes less steep than on the red part, thus decreasing $S_w$. Although our simulated gas density and velocity field are far more complex that those underlying the Neufeld solution, the above interpretation continues to hold. For this reason we find small or even negative values of $S_w$. 
As photoionization proceeds, $N_{\rm{HI}}$  drops below $10^{18}\rm{cm^{-2}}$, the peak position gets closer to the line center, and the profile becomes highly skewed as typically observed in LAEs.
Most cases have $S_w < 0$ initially, later increasing as $\NHI$ decreases. 

\subsection{Surface brightness evolution}

Star-forming galaxies are more compact at high $z$: typically, at $z\gsim6$ they have a half-light radius of 1 kpc \citep{Bouw04}. Therefore the Ly$\alpha$ emitting area is smaller than the angular resolution ( $\simeq 0.8"$) of narrow band images and the SB radial profile from the brightest pixel approximates  a Gaussian point spread function (PSF) of s.d. $\sigma=\rm{seeing}/2.35$. If Ly$\alpha$ emission is powered by X-rays, however, it can be spatially more extended, as already stated.

\begin{figure*}
 \includegraphics[width=55mm]{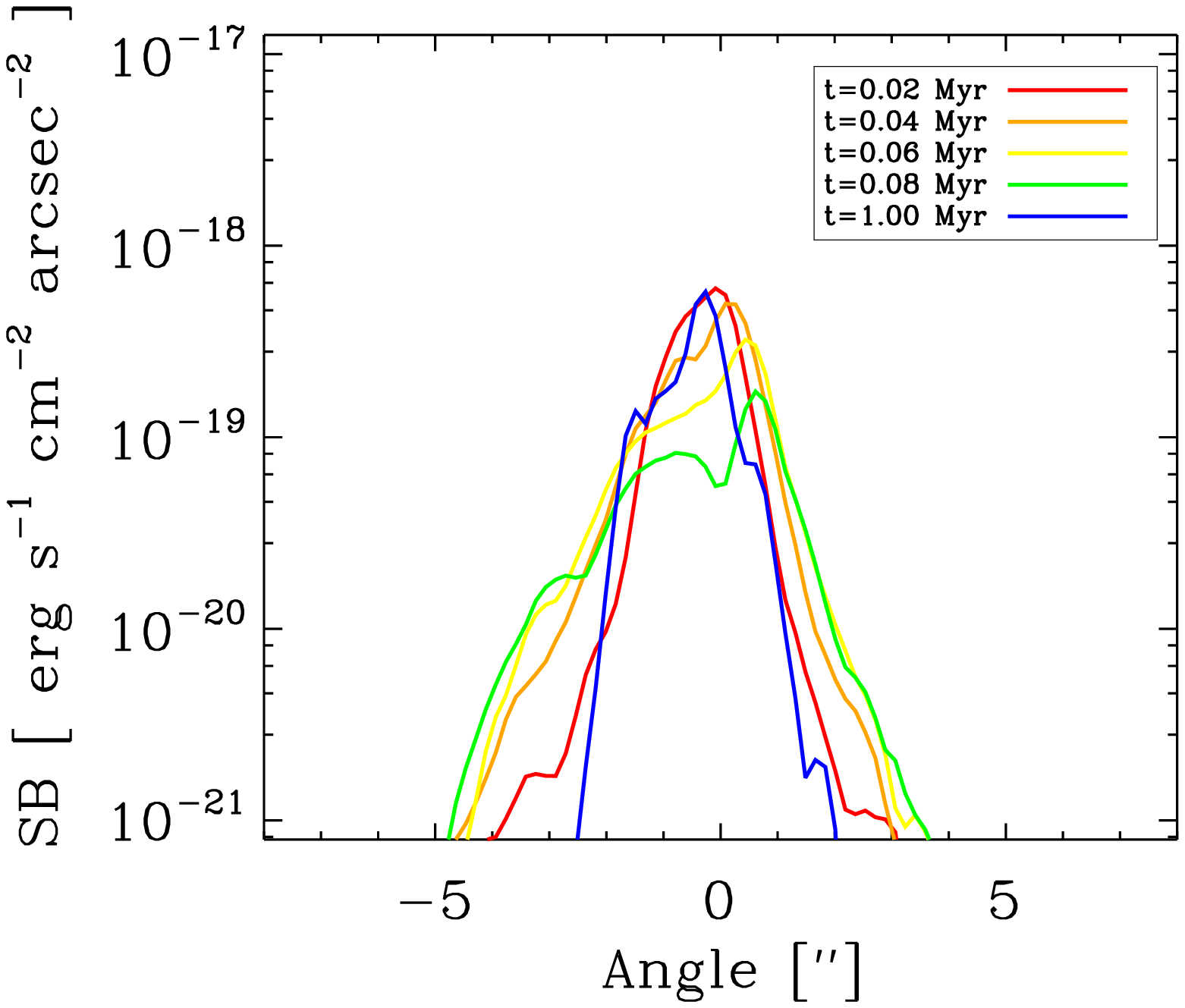}
 \includegraphics[width=55mm]{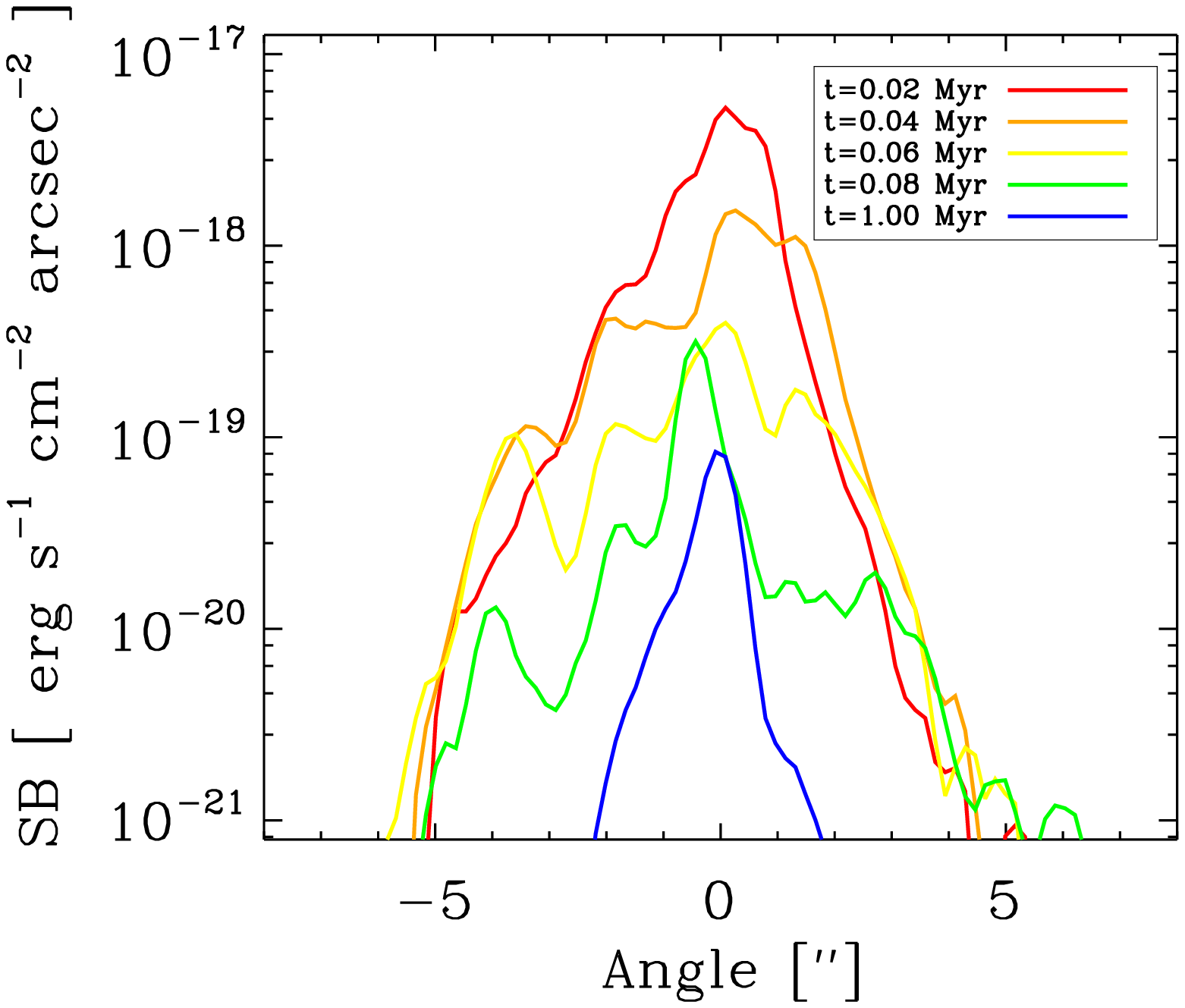}
 \includegraphics[width=55mm]{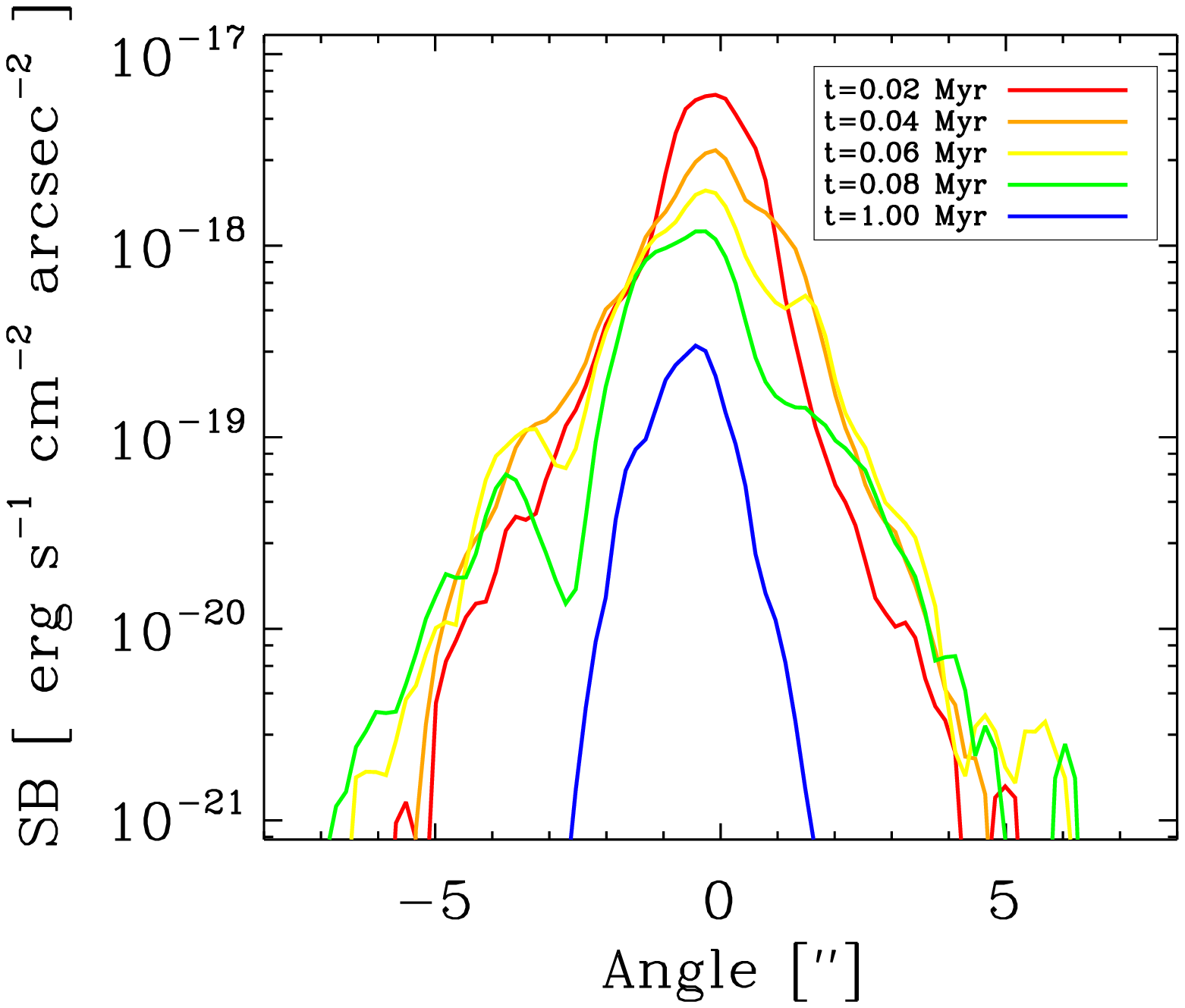}
 \caption{SB profile evolution through the brightest pixel of the maps in Fig. \ref{panel} at $t=0.02,\,0.04,\,0.06,\,0.08,\,1.00$ Myr}
 \label{SB}
\end{figure*}

Fig. \ref{SB} shows the simulated SB profiles for the three SED cases. For the starburst, the SB profile is initially relatively broad  (FWHM $\simeq1.2"$) since Ly$\alpha$ photons are very scattered in space; later on, the FWHM decreases and approaches the resolution scale expected for a Ly$\alpha$ point source; even in the early evolutionary phases, though, the FWHM never exceeds $1.4"$. On the contrary, for both AGN cases several pixels are brighter than the detection limit when the Ly$\alpha$ photon production is dominated by X-rays (Fig. \ref{SB}). The width is $\simgt 1.5"$ for the Compton-thin AGN case.  After 1 Myr, though,  the flux fades below the detection limit and the profile becomes narrower since the total Ly$\alpha$ luminosity decreases.

\subsection{Identification criterion}

\begin{figure}
 \includegraphics[width=80mm]{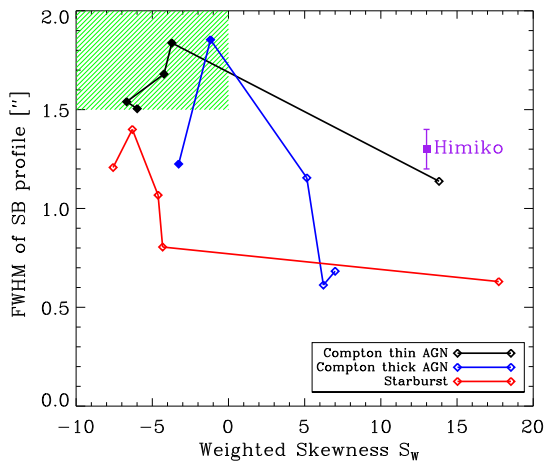}
 \caption{FWHM of SB profile vs. skewness. From left to right, each model has 5 data points at $t=0.02,\,0.04,\,0.06,\,0.08,\,1.00$ Myr.  The shaded area denotes AGN-powered LAE identification criterion $S_w<0$ and FWHM $\ge1.5"$, 
}
 \label{two_critere}
\end{figure}

From the previous results we can build a robust criterion to identify AGN-powered LAEs, as shown in Fig. \ref{two_critere}
where we plot the evolutionary tracks of the three SEDs in the FWHM vs. $S_w$ plane.   
As $S_w$ is anti-correlated with $\NHI$, it increases with time for all models. Thus $S_w<0$ alone cannot be used to uniquely identify AGN-powered LAEs. However, the starburst FWHM cannot exceed $1.4"$, independently of the assumed SFR as long as 
it is sufficient to allow detection of the source as a LAE ($L_\alpha > 10^{42.2}\,\rm{erg\,s^{-1}}$). Therefore, the criterion defined by $S_w<0$ and $FWHM\ge1.5"$, can be safely used to identify the class of LAEs powered by AGN, if they exist. 

\section{Conclusions}

The presence of nuclear black holes, progenitors of the super-massive ones discovered at $z=6-7$, in high redshift
galaxies is expected on several grounds. Yet direct detection of their X-ray emission has so far proven very challenging,
particularly at $z>6$. In addition some of these AGNs can be heavily obscured and show no broad emission lines.
 Thus it would be useful to be able to identify AGN-powered LAEs purely from their observed 
\Lya properties. 

We have explored this possibility using RT cosmological simulations of one of the most luminous LAE known, Himiko. 
From the simulated \Lya luminosities, line profile and surface brightness we have been able to isolate key differences among
three plausible SEDs of the central source (starburst, Compton-thick or Compton-thin AGN types) with a fixed 
photoionization rate $Q=3\times 10^{54}\,\rm{s^{-1}}$.

From the results we have built a robust criterion to identify AGN-powered LAEs. We find that these sources should have (a): \textit{negative} line weighted skewness, $S_w <0$, and (b) surface brightness profiles FWHM $\ge1.5"$; this parameter space cannot be populated by starburst LAEs. Thus, LAEs satisfying this criterion would be strong candidates for the presence of a hidden AGN powering their luminosity.  Note that this criterion purely depends on the observed properties of the \Lya line and does not require additional information from, say, UV continuum or X-ray data. According to such criterion, Himiko, due to 
its high skewness, $S_w=13.2$ and somewhat too narrow SB profile (Fig.\ref{two_critere}), is not predicted to host an AGN.

Clearly the success of the proposed strategy depends on the sensitivity of the observations. A sufficiently deep SB map will 
allow to accurately measure the SB radial profile over a wide range of distances where the differences between starburst and AGNs become more appreciable (see Fig. \ref{SB}). In this sense, the advent of the new generation of instruments, as e.g. MUSE, 
will be crucial.

\bibliographystyle{apj}
\bibliography{ref}

\end{document}